\documentclass[10pt]{article}
\usepackage{amsmath}
\usepackage{amssymb}
\usepackage{graphicx}
\usepackage{epstopdf}
\usepackage[mathscr]{eucal}

\usepackage[latin5]{inputenc}

\usepackage{url}
\usepackage{authblk}

\numberwithin{equation}{section}

\newcommand{\be}{\begin{equation}}
\newcommand{\ee}{\end{equation}}

\begin{document}

\clearpage
\thispagestyle{empty}

\title{The energy level structure of a variety of one-dimensional confining potentials and the effects of a local singular perturbation}

\author[1,2]{M.L. Glasser\thanks{laryg@tds.net}}
\author[1]{L.M. Nieto\thanks{luismi@fta.uva.es}}
\affil[1]{Dept. de F\'{\i}sica Te\'{o}rica, At\'{o}mica y \'{O}ptica, Univ. de Valladolid, 47011 Valladolid, Spain}
\affil[2]{Department of Physics, Clarkson University, Potsdam, NY
13699, USA}

\date{\today}

\maketitle

\begin{abstract}
Motivated by current interest in quantum confinement potentials, especially with respect to the Stark spectroscopy of new types of quantum wells, we examine several novel one-dimensional singular oscillators. A Green function method is applied, the construction of the necessary resolvents is reviewed and several new ones are introduced. In addition, previous work on the singular harmonic oscillator model, introduced by Avakian et al. is reproduced to verify the method and results. A novel features is the determination of the spectra of asymmetric hybrid linear and quadratic potentials. As in previous work, the singular perturbations are modeled by delta functions.
\end{abstract}

\maketitle

\section{Introduction}

The bound state spectra in confining potentials, especially of linear and quadratic form, has been a concern in quark confinement, for example, for some time \cite{LEP}. The effect on these spectra due to local changes in the potential at specific points is also of concern and led Avakian et al. \cite{Avakian} (see also \cite{At,Vian}) to introduce the singular harmonic oscillator model
$V(x)=a x^2+b\delta(x)$ which, along with generalizations, has accumulated an extensive literature. Of particular interest is the extensive body of fundamental work on these systems by S. Fassari et al. \cite{FI1,FI2,FI3,FR}, and that summarized in the book of Albeverio et al. \cite{albeverio} (see also \cite{Bel}) to which we refer the reader for  further references.
More recently, these models have arisen in the study of semi-conductor quantum wells where, since the 1980's, it has been possible to engineer these entities with increasingly sophisticated properties. Quantum wells having parabolic confinement in applied electric fields are fundamental for a growing  optical device industry. For a survey  of the basic experimental and theoretical work in this area see, e.g. \cite{Denev}.
In the pioneering work these systems could be treated successfully  as  singularly decorated square wells and a Green function matching technique was developed  providing an accurate account of the electro-luminescence  of Ga-As based quantum wells \cite{Barrio}.
In this note we present a first attempt to extend this work to the  quantum-confined Stark effect in  low symmetry and anisotropic structures now being explored. For details and references see \cite{Porubaev}, e.g.

Our aim in this note is to reproduce the basic energy level calculations for the harmonic case by  the  expeditious Green function method derived in \cite{comp} and then to go on to examine a number of new systems, including the important linear oscillator $V(x)=a|x|+b\delta(x-q)$ which, surprisingly, has not been studied in detail. For completeness, we begin by deriving the basic Green functions.

For a one-dimensional simple harmonic oscillator having frequency $\omega$ 
\be
\begin{array}{l}
\displaystyle
{\cal H}_{ho}= -\frac{\hbar^2}{2m}\frac{d^2}{dx^2}+ \frac{m\omega^2}2  x^2,\qquad  
E^{ho}_n=\hbar\omega\left(n+\frac12\right),
\\ [2ex]
\displaystyle
\psi^{ho}_n(x)=\frac1{\sqrt{2^n n!}} \left( \frac{m\omega}{\pi\hbar} \right)^{1/4}
e^{-{m\omega x^2}/(2\hbar)} H_n\left(\sqrt{\frac{m\omega}{\hbar}} x\right),
\end{array}
\label{harmonicoscillator}
\ee
the causal propagator
\begin{eqnarray}
K_{ho}(x,x';t)
&=&\sum_{n=0}^{\infty} \psi^{ho}_n(x)(\psi^{ho}_n(x'))^*e^{-iE^{ho}_nt/\hbar}
\nonumber \\
&=&\sqrt{\frac{m \omega}{2\pi i\hbar \sin\omega t}}\exp\left[\frac{-m\omega}{2i\hbar}\Bigl[(x^2+x'^2)\cot\omega t-2xx'\csc\omega t\Bigr]\right] \theta(t),
\label{1dhopropagator}
\end{eqnarray}
where $\theta(t)$ represents the Heaviside step function, is derived in many elementary texts, e.g.  \cite{Feynman,Manoukian} and plays an important role in problems concerning electrons in magnetic fields, for example. However, its time-Fourier transform, the equally useful energy dependent Green function (Schr\"odinger resolvent)
\be\label{definitionofgreen}
G_{ho}(x,x';E)=\frac{i}{\hbar} \int_0^{\infty}e^{iE^+t/\hbar} K_{ho}(x,x';t)dt
\ee
is less well-known, although it seems to have first appeared on p.~74 of the classical book of Titchmarsh  \cite{Titchmarsh}. In the case of the three-dimensional oscillator \eqref{definitionofgreen} was evaluated explicitly  by Krebtukov and Macek \cite{Khrebtukov}. In the next section we  present an independent derivation of $G_{ho}(x,x';E)$ in one dimension in terms of the Parabolic Cylinder Function $D_{\nu}(x)$ (see Chap.~8 of \cite{Erdelyi}), by obtaining an integral representation for the  Hermite polynomials resolvent series. We study two applications of this result: the case of adding a linear potential to the harmonic oscillator and the case of the hybrid asymetric oscillator proposed in \cite{comp}; in both cases the spectrum is analyzed in detail. In section 3 we obtain the Green function for the linear confining potential  $V(x)=\alpha^3 |x|$, in terms of Airy functions. The applications to an asymmetric linear potential and to  a couple of different combinations of oscillator and linear potentials are studied.
 In Section~4 we revisit the $\delta$-decorated harmonic well problem \cite{Avakian}, and we analyze the same type of ``decoration'' for the linear confining potential  previously mentioned. The paper ends with a summary and discussion.

\section{Calculation of the harmonic oscillator Green function}

In order to determine the Green function of a system uniquely, the boundary conditions must be ``built in"  to the solution of the Green function equation. In the case of (1.3) this is achieved by adjusting the integration contour. For the three dimensional oscillator, this was achieved \cite{Khrebtukov} by assigning the energy an infinitesimal imaginary part. The subsequent integration was then carried out by taking advantage of the rotational symmetry of the system. In the one dimensional case this is not an option and we must proceed differently.

From the familiar eigenstates $\psi^{ho}_n(x)$ and eigenergies $E^{ho}_n$ of the harmonic oscillator \eqref{harmonicoscillator}, as for any Sturm-Liouville system \cite{Titchmarsh}, we form the Green function (resolvent)  as a series, where the boundary conditions are automatically incorporated into the eigenfunctions,
\be
G_{ho}(x,x';E)=\sum_{n=0}^{\infty} \frac{\psi^{ho}_n(x)(\psi^{ho}_n(x'))^*}{E^{ho}_n-E}=
\sqrt{\frac{m\omega}{\pi\hbar}}e^{-m\omega(x^2+x'^2)/(2\hbar)}\sum_{n=0}^{\infty}\frac{H_n(\sqrt{\frac{m\omega}{\hbar}}x)H_n(\sqrt{\frac{m\omega}{\hbar}}x')}{2^nn![\hbar\omega(n+\frac{1}{2})-E]}.
\label{(initial3)}
\ee
As it is well known, this Green function satisfies the general equation\footnote{In operator form $(H-E)G=I=G(H-E)$.}
\be
\left[ -\frac{\hbar^2}{2m}\frac{\partial^2}{\partial x^2}+V(x)-E \right] G(x,x';E)=\delta(x-x')
\label{generaldefinitionofgreenfunction}
\ee
for the oscillator potential $V(x)=m\omega^2 x^2/2$. Sometimes another dimensionless version of the Green function is used, let us call it ${\widetilde G}(x,x'; \epsilon=E/(\hbar\omega))$, related to $G(x,x';E)$ as
\be
{\widetilde G}(x,x'; \epsilon)= -\frac{\hbar^2}{2m} G(x,x';E), \quad 
\left[\partial_x^2+ \mu^2\epsilon- \frac{\mu^2}{\hbar\omega} V(x) \right] {\widetilde G} (x,x'; \epsilon)=\delta(x-x'),  \quad \mu=\sqrt{\frac{2m\omega}{\hbar}} .
\label{greentilde}
\ee
In this particular case, in order to simplify the calculations, instead of working with the function \eqref{(initial3)} let us set up the equivalent dimensionless system obtained by introducing the dimensionless quantities $y=\mu x/\sqrt{2}$ and  $\psi(x)= \sqrt{\mu/ \sqrt{2}}\, \phi(y)$:
\be\label{adimensionalHO}
\widehat{\cal H}=\frac{2 {\cal H}_{ho}}{\hbar \omega}=
-\frac{d^2}{d y^2}+ y^2,\quad
\phi_n(y)=\frac1{\sqrt{2^n n!\sqrt{\pi}}} \ 
e^{-y^2/2} H_n\left(y\right),
\quad \lambda_n=2n+1.
\ee
Then,
\be
G_{ho}(x,x';E)= \sqrt{\frac{m}{\omega\hbar^3}}\
\widehat{G}\left(y,y'; \epsilon \right), \quad 
\widehat{G} \left( y,y'; \epsilon \right)=
\frac{e^{-(y^2+y'^2)/2}}{\sqrt{\pi}} \sum_{n=0}^{\infty}\frac{H_n(y)H_n(y')}{2^nn! \left( n+\frac{1}{2}-\epsilon \right)}.
\label{(3)}
\ee
Next, consider the sum
\be\label{(333)}
S(z,w;s)=\sum_{n=0}^{\infty}\frac{H_n(z)H_n(w)}{2^nn!(n+s)}= \sqrt{\pi}e^{(z^2+w^2)/2}\ \widehat{G} \left(z,w;\frac{1}{2}-s\right),
\ee
which is a meromorphic function of $s$ having non-positive integer poles. Thus if we obtain its value in any singularity-free region of the complex plane, analytic continuation can be invoked to determine its value elsewhere. Let us, therefore, begin by assuming that $\text{Re}[s]>2$ and use the representation $\int_0^1u^{n+s-1}du=(n+s)^{-1}$ to get
\be
S(z,w;s)=\int_0^1du\frac{u^{s-1}}{\sqrt{1-u^2}}\exp\left[\frac{2z w u-(z^2+w^2)u^2}{1-u^2}\right], \label{555}
\ee
where Mehler's formula (see Sec.~10.13(22) of \cite{Erdelyi}) has been used to sum the  series. Next, by means of the change of variables $t=u^2/(1-u^2)$ equation \eqref{555} becomes the Laplace  representation
\be
S(z,w;s)=\frac{1}{2}\int_0^{\infty}t^{s/2-1}(t+1)^{-s/2-1/2}e^{-(z^2+w^2)t}e^{2z w\sqrt{t(t+1)}}dt.\label{666}
\ee
Now, according to \cite{Buchholtz}, for $\nu>0$ and $a>b>0$,
$$
\int_0^{\infty} \frac{t^{\nu-1}}{(1+t)^{\nu+1/2}}\ e^{b\sqrt{t(1+t)}-a t}dt =2 e^{a/2} \ \Gamma(2\nu) \ D_{-2\nu} \left[ \sqrt{a+\sqrt{a^2-b^2}}\right]   D_{-2\nu} \left[-\sqrt{a-\sqrt{a^2-b^2}} \right].\label{lap}
$$
After a simple calculation we obtain 
\begin{eqnarray}
S(z,w;s)= e^{(z^2+w^2)/2}\ \Gamma(s)\, D_{-s}(\sqrt{2}\, z_>)\ D_{-s} (-\sqrt{2}\, z_<) ,
\label{(7)}  
\end{eqnarray}
where $z_<\,(z_>)$ represent the  smaller (larger) of $z,w$.
The expression \eqref{(7)} is analytic  throughout the $s$-plane, except at the poles of the gamma function. This expression for $S$ does not seem to appear in the literature on Hermite functions.

We can now replace in \eqref{(7)} $z, w,s$ in terms of $x, x',E$ and take into account \eqref{(3)}--\eqref{(333)} to obtain
\be
G_{ho}(x,x';E)= 
\sqrt{\frac{m}{\pi\omega\hbar^{3}}}\
\Gamma\left(\frac{1}{2}-\epsilon\right)
\ 
D_{\epsilon-\frac{1}{2}} \left(\mu\, x_>\right)
\
D_{\epsilon-\frac{1}{2}} \left( -\mu\, x_<\right),
\label{(8)}
\ee
where now $x_<\,(x_>)$ represent the  smaller (larger) of $x,x'$, and the physical parameters $\mu$, $\epsilon$ are given in \eqref{greentilde}.
The poles of \eqref{(8)}, and therefore the bound states, are precisely the poles $-n$ of the Gamma function, giving back the result \eqref{harmonicoscillator}.

\begin{figure}[h]
\centering
\includegraphics[width=0.4\textwidth]{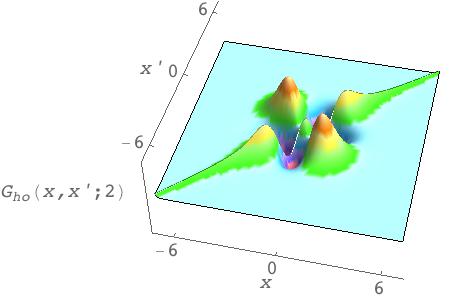}
\qquad
\includegraphics[width=0.4\textwidth]{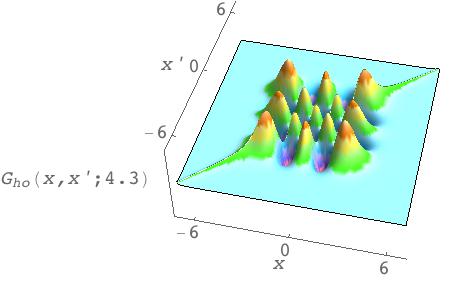}
\caption{\small Green function  \eqref{(8)} in units such that $\omega=\hbar=m=1$ for two energy values: on the left $G_{ho}(x,x';E=2)$ and on the right $G_{ho}(x,x';E=4.3)$.
\label{fig1}}
\end{figure}

In Figure~\ref{fig1} we show plots of $G_{ho}(x,x';E)$ as a function of $x$ and $x'$, for a couple of values of the energy. One can appreciate a lot of symmetry in these plots due to the fact that   the eigenfunctions of the oscillator entering the Green function \eqref{(initial3)} are real, and either even or odd, implying that $G_{ho}(x,x';E)=G_{ho}(x',x;E)$ and $G_{ho}(x,x';E)=G_{ho}(-x',-x; E)$, as is clearly seen in the figures.

\subsection{Harmonic oscillator plus uniform electric field}

Let us analyze first the case of a uniform electric field, given by the linear potential $\alpha^3\, x$, added to the original oscillator potential. By completing the square in the Schr\"odinger equation, it is easily seen that the system is again a simple harmonic oscillator, but with shifted coordinate $x_\alpha = x+\varphi$ and shifted energy $E^\alpha = \hbar\omega(\epsilon-(\mu\varphi/2)^2)$: 
\be
{\cal H}_{ho,\alpha}= -\frac{\hbar^2}{2m}\frac{d^2}{dx^2}+ \frac12 m\omega^2 x^2+\alpha^3 x=
\left[ -\frac{\hbar^2}{2m}\frac{d^2}{d x_\alpha^2} + \frac12 m\omega^2 x_\alpha^2\right] -\hbar\omega \left(\frac{\mu\varphi}{2}\right)^2, 
\label{harmonicoscillator+Fx}
\ee
where
\be 
\epsilon=\frac{E}{\hbar\omega}, \quad \varphi= \frac{\alpha^3}{m\omega^2}, \quad   \mu=\sqrt{\frac{2m\omega}{\hbar}}, \quad
\sigma=\epsilon+\left(\frac{\mu\varphi}2\right)^2.
\label{physicalparameters}
\ee
The Green function for this system is then simply
\begin{eqnarray}\label{(10)}
G_{ho,\alpha}(x,x';E)  \!\!& \!\!= \!\! &\!\!  G_{ho} \left(x+\varphi ,x'+\varphi ; E+\hbar\omega(\mu\varphi/2)^2 \right)
\\ [1ex]
 \!\!& \!\!= \!\! &\!\! \sqrt{\frac{m}{\pi\omega\hbar^{3}}}\ 
\Gamma\left(\frac{1}{2}-\epsilon-\frac{\mu^2\varphi^2}{4} \right)
 D_{\sigma-\frac12} \bigl(\mu (x_> +\varphi ) \bigr)\
D_{\sigma-\frac12} \bigl( -\mu (x_< +\varphi) \bigr),
\nonumber
\end{eqnarray}
where all the physical parameters are given in \eqref{physicalparameters}.

\begin{figure}[h]
\centering
\includegraphics[width=0.4\textwidth]{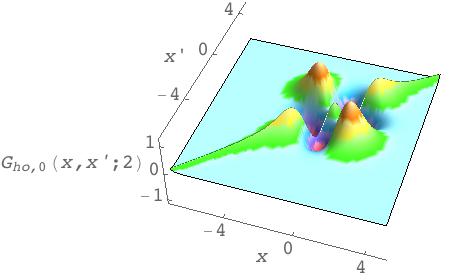} 
\qquad
\includegraphics[width=0.4\textwidth]{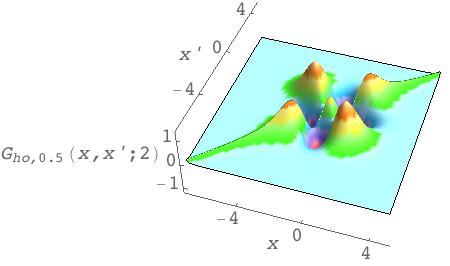}
\\
\includegraphics[width=0.4\textwidth]{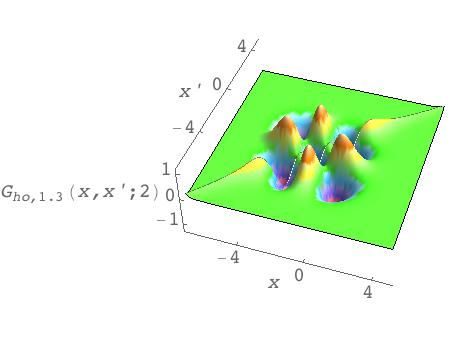} 
\qquad
\includegraphics[width=0.4\textwidth]{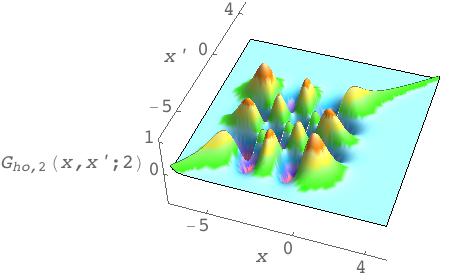}
\caption{\small Green function  \eqref{(10)} in units such that $\omega=\hbar=m=1$ for a fixed value of the energy ($E=2$) and four different values of the electric field: from left to right, and from top to bottom, $\alpha^3=0, 0.5, 1.3,$ and $2$.
\label{greenoscillpluslinear}}
\end{figure}

A plot of the Green function $G_\alpha(x,x';E)$ is shown in Figure~\ref{greenoscillpluslinear} for $E=2$ and four different values of the electric field. The symmetry that appeared in $G_{ho}(x,x';E)$ is still present in $G_\alpha(x,x';E)$ because they are related through \eqref{(10)}; the symmetry center is no longer the origin but the point $(-\varphi,-\varphi)$ on the $(x,x')$--plane.
The poles of \eqref{(10)}, and therefore the bound states,  are
\be
E^\alpha_n=\hbar\omega \left(n+\frac12 -\left(\frac{\mu\varphi}{2}\right)^2   \right), \quad n=0,1,2,\dots
\ee
From these expressions it is quite obvious that in the limit $\alpha,\varphi\to 0$ we recover the bound states energies and the Green function of the simple harmonic oscillator.

\subsection{Asymmetric harmonic oscillator potential}

\begin{figure}[h]
\centering
\includegraphics[width=0.33\textwidth]{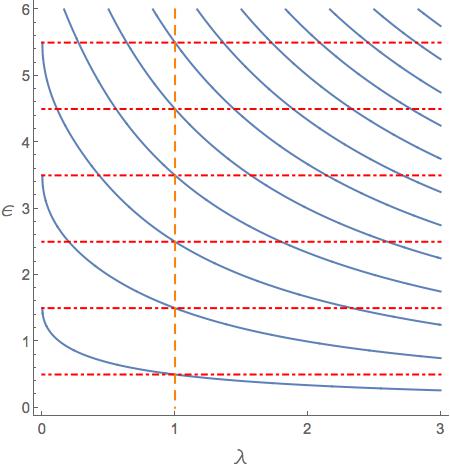} \qquad\qquad 
\includegraphics[width=0.33\textwidth]{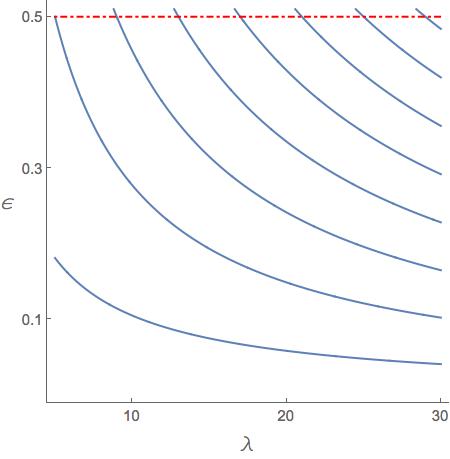} 
\caption{\small Bound states of the asymmetric oscillator $\epsilon_n=E_n/(\hbar \omega_1)$ vs $\lambda=\omega_1/\omega_2$ (solid curves), for small values of $\lambda$ (on the left) and bigger values of $\lambda$ (on the right). The harmonic oscillator energy eigenvalues are indicated with dot-dashed horizontal red lines, and the harmonic oscillator case with a vertical dashed orange line.
\label{asymetricoscillatorenergies}}
\end{figure}

As a second application, consider the asymmetric oscillator potential
\be
V(x)=\left\{\begin{array}{cc}
\frac{1}{2} m\omega_1^2 x^2,&  x\leq 0, \\ [2ex]
\frac{1}{2} m\omega_2^2 x^2,&  x\geq 0,
\end{array}
\right.
\label{asymetricoscillator}
\ee
and let $G_j(x,x';E)$ denote the Green function $G_{ho}(x,x';E)$ given in \eqref{(8)} evaluated for $\omega=\omega_j$. Then the asymmetric oscillator states, which do not coincide with $(n+1/2)\hbar \omega_j$, $j=1,2$, are given by \cite{comp}
\be
G_1(0,0;E)+G_2(0,0;E)=0.
\label{(3.5)}
\ee
In terms of $\epsilon=E/(\hbar \omega_1)$ and $\lambda=\omega_1/\omega_2$, this is
\be
\frac{1}{\displaystyle
 \Gamma\left(\frac{1}{4}-\lambda \frac{\epsilon}2 \right)
 \Gamma\left(\frac{3}{4}-\frac{\epsilon}2 \right)}
+ \frac{\displaystyle\lambda^{1/2}}{\displaystyle
\Gamma\left(\frac{1}{4}-\frac{\epsilon}2 \right)
\Gamma\left(\frac{3}{4}- \lambda \frac{\epsilon}2 \right)}=0 .
\label{(3.6)}
\ee
Note that, from equation~\eqref{(3.6)}, $\epsilon(\lambda)=\epsilon(1/\lambda)/\lambda$, so we need only examine $0<\lambda<1$. For $\lambda=1$ (the symmetric harmonic oscillator $\omega_1=\omega_2$) there are clearly no new eigenvalues, apart from the already known \eqref{harmonicoscillator}. 
A plot of the eigenvalues $\epsilon=E/(\hbar \omega_1)$ as a function of $\lambda=\omega_1/\omega_2$  is given in Figure~\ref{asymetricoscillatorenergies}, where the harmonic oscillator eigenvalues are also indicated by the horizontal red lines. Note that in this plot it is quite clear that when $\lambda=1$, that is, for the symmetric oscillator, the correct eigenvalues are recovered.

For $\lambda\to 0$, that is, $\omega_2\to\infty$ and $\omega_1$ fixed, the potential \eqref{asymetricoscillator} physically corresponds to an infinite barrier for $x>0$ and a semioscillator of frequency $\omega_1$ for $x<0$; in this case the spectrum is well known to be $(2n+1+\frac12)\hbar\omega_1$, something that is also clearly seen in the left part of Figure~\ref{asymetricoscillatorenergies}: note that the states of the simple harmonic oscillator ($\lambda=1$) evolve as $\lambda\to 0$ in such a way that
\be
\lim_{\lambda\to 0}\epsilon_0(\lambda)=\frac32, \quad \lim_{\lambda\to 0}\epsilon_1(\lambda)=\frac72, \quad\dots \quad \lim_{\lambda\to 0}\epsilon_n(\lambda)=2n+1+\frac12,  \quad n=0,1, \dots,
\ee
and then the  spectrum $\epsilon_n(\lambda=1)=(n+\frac12)$ is rescaled in a continuous way to  become $\epsilon_n(\lambda=0)=(2n+1+\frac12)$ in the limit $\lambda\to 0$ or $\omega_2\to\infty$.

The case $\lambda\to \infty$ can be imagined as $\omega_2\to 0$ and $\omega_1$ fixed, and corresponds to a semi-oscillator of frequency $\omega_1$ for $x\leq 0$ and $V=0$ for $x\geq 0$, a situation without bound states, a fact that is suggested in the right part of Figure~\ref{asymetricoscillatorenergies} by the fact that 
\be
\lim_{\lambda\to \infty}\epsilon_n(\lambda)=0, \quad n=0,1, \dots
\ee

We should mention that, although in the present application we have concentrated in obtaining a closed transcendental expression for the bound states, it is also possible to obtain the Green function for the asymmetric potential. Nevertheless, the task is quite cumbersome and we reserve it for the future.

\section{Calculation of the Green function for $V(x)=\alpha^3\, |x|$}

As a second example, let us consider now the Schr\"odinger equation for a linear potential of the form
\be
-\frac{\hbar^2}{2m}\frac{d^2 \psi(x)}{dx^2}+\left( \alpha^3\, |x| -E \right) \psi(x)=0.
\label{absolutevaluepotential}
\ee
A related problem was considered in Ref.~\cite{GGV}. 
It is well known that equation \eqref{absolutevaluepotential} can be solved in terms of Airy functions. The general solution is
\be
\psi(x)=\left\{\begin{array}{ll}
C_1\, {\rm Ai}(-\zeta x-\varrho) + C_2\, {\rm Bi}(-\zeta x-\varrho) ,&  x\leq 0, \\ [2ex]
C_1\, {\rm Ai}(\zeta x-\varrho) + C_2\, {\rm Bi}(\zeta x-\varrho)   ,&  x\geq 0,
\end{array}
\right.
\label{generalsolutionsairy}
\ee
where
\be
\varrho=\frac{E}{\alpha^2}  \left( \frac{2m}{\hbar^2} \right)^{1/3} ,  \qquad \zeta=\alpha \left( \frac{2m}{\hbar^2} \right)^{1/3} .
\label{parameters_linearcase}
\ee
As we require $\psi(\pm\infty)=0$, $C_2=0$.
Now consider the differential equation defining the Green function $G_\alpha(x,x';E)$ for equation \eqref{absolutevaluepotential}, similar to \eqref{generaldefinitionofgreenfunction}, or ${\widetilde G}_\zeta(x,x'; \varrho)=-\hbar^2 G_\alpha(x,x';E)/(2m)$, similar to 
\eqref{greentilde}:
\be
\bigl[\partial_x^2+\zeta^2\varrho- \zeta^3\, |x|\bigr]{\widetilde G}_\zeta(x,x'; \varrho)=\delta(x-x').
\label{partialdifeqnforgreenabsx}
\ee
Clearly ${\widetilde G}_\zeta(x,x'; \varrho)$ must be continuous for $x=x'$ and ${\widetilde G}_\zeta(x,x'; \varrho)= {\widetilde G}_\zeta(x',x; \varrho)$. Hence, the solution must have the form
\be
{\widetilde G}_\zeta(x,x'; \varrho)=\left\{\begin{array}{cc}
C\ {\rm Ai}(\zeta x-\varrho)\ {\rm Ai}(-\zeta x'-\varrho), &x\geq x', \\  [2ex]
C\ {\rm Ai}(-\zeta x-\varrho)\ {\rm Ai}(\zeta x' -\varrho), &x\leq x'.
\end{array}\right.
\ee
But, there is also the jump condition in the first derivative. In order to generate the $\delta$-function on the RHS of \eqref{partialdifeqnforgreenabsx}, the partial derivative $\partial_x {\widetilde G}_\zeta(x,x'; \varrho)$ must be discontinuous:
\be
\lim_{\epsilon\to 0} \left(  
\left. \partial_x {\widetilde G}_\zeta(x,x'; \varrho) \right|_{x=x'+\epsilon}-  
\left. \partial_x {\widetilde G}_\zeta(x,x'; \varrho) \right|_{x=x'-\epsilon}
\right)=1.
\ee
In the present case we must have
\be
C\, \zeta\, \bigl[ {\rm Ai}(\zeta x'-\varrho)\ 
{\rm Ai}'(-\zeta x'-\varrho)+{\rm Ai}(-\zeta x'-\varrho)\ {\rm Ai}'(\zeta x'-\varrho) \bigr]=1.
\label{thejumponthefirstderivativeofGreen}
\ee
Since the quantity in \eqref{thejumponthefirstderivativeofGreen} is essentially the Wronskian, it is independent of $x'$, so we can set $x'=0$ to get
\be
G_\alpha(x,x';E)=- \left( \frac{2m}{\hbar^2} \right)^{2/3}  \frac1{2\alpha}  \frac{{\rm Ai}(\zeta x_>-\varrho)\ 
{\rm Ai}(-\zeta x_<-\varrho)}{{\rm Ai}(-\varrho) \ {\rm Ai}'(-\varrho)},
\label{finalgreenabsgreen}
\ee
where, as in the previous section, $x_<\,(x_>)$ represent the  smaller (larger) of $x,x'$, and the physical parameters $\varrho$ and $\zeta$ are given in \eqref{parameters_linearcase}. 
The poles of \eqref{finalgreenabsgreen} are the bound states of this problem: basically the alternating zeros of ${\rm Ai}'(z)$ (even states) and ${\rm Ai}(z)$ (odd states). Formulas and list of numerical values of these zeros can be found in many references, for example \cite{NIST}

\begin{figure}[h]
\centering
\includegraphics[width=0.4\textwidth]{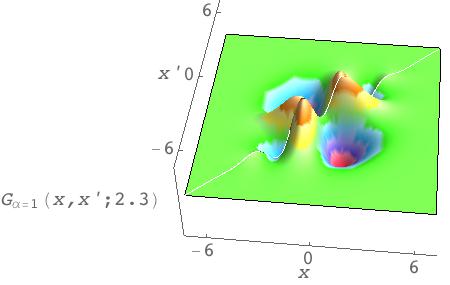} \qquad
\includegraphics[width=0.4\textwidth]{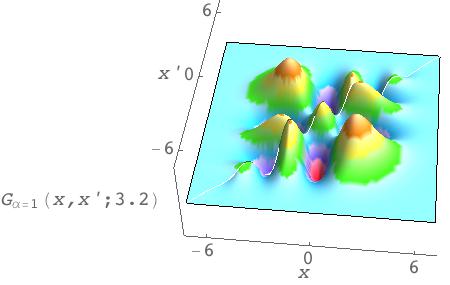}
\caption{\small Green function of equation \eqref{finalgreenabsgreen} for $\hbar^2=2m$ and $\alpha=1$, that is, $\varrho=E$ and $\zeta=1$, for two energy values: on the left $G_{\alpha=1}(x,x';E=2.3)$ and on the right $G_{\alpha=1}(x,x';E=3.2
)$.
\label{figura4}}
\end{figure}

In Figure~\ref{figura4} we show plots of $G_{\alpha}(x,x';E)$ as a function of $x$ and $x'$, for a couple of values of the energy. As in the harmonic oscillator, a lot of symmetry is also present in these plots, a fact which is due to the symmetry of the potential, implying that $G_{\alpha}(x,x';E)=G_{\alpha}(x',x;E)=G_{\alpha}(-x',-x; E)$, as it is clearly seen in the figures.

\subsection{Asymmetric linear potential}

\begin{figure}[h]
\centering
\includegraphics[width=0.33\textwidth]{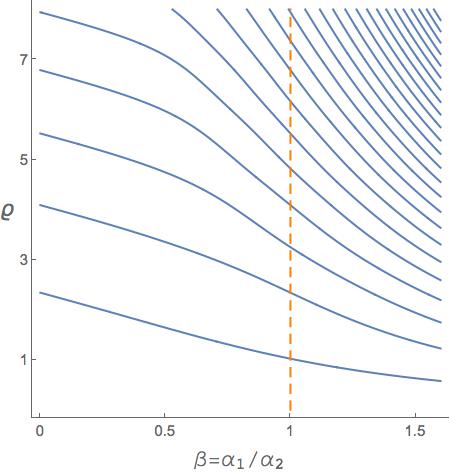} 
\caption{\small Bound states of the asymmetric absolute value potential 
\eqref{asymetriclinear} given implicitly by \eqref{asymmetricVpotentialstateseqn}: the physical parameter giving the energy, $\varrho=( 2m/\hbar^2)^{1/3} E/{\alpha_1^2}$, is represented as a function of the frequency quotient $\beta=\alpha_1/\alpha_2$.
\label{asymmetricVpotentialstatesplot}}
\end{figure}

As an application of the result we have obtained in \eqref{finalgreenabsgreen}, let us consider the asymmetric linear potential
\be
V(x)=\left\{\begin{array}{cc}
-\alpha_1^3 x,&  x\leq 0, \\ [1ex]
\alpha_2^3 x^2,&  x\geq 0,
\end{array}
\right.
\label{asymetriclinear}
\ee
and let $G_{\alpha_j}(x,x';E)$ denote the Green function $G_{\alpha}(x,x';E)$ given in \eqref{finalgreenabsgreen} evaluated for $\alpha=\alpha_j$. Then the asymmetric linear states, are given by the general result \eqref{(3.5)} (taken from \cite{comp}):
\be
G_{\alpha_1}(0,0;E)+G_{\alpha_2}(0,0;E)=0.
\label{eigenvaluesasymmetriclinear}
\ee
In terms of $\varrho=( 2m/\hbar^2)^{1/3} E/{\alpha_1^2} $ and $\beta=\alpha_1/\alpha_2$, this can be written as
\be
-\frac{{\rm Ai}(-\varrho)/{\rm Ai}'(-\varrho)}{ {\rm Ai}(-\varrho \beta^2)/{\rm Ai}'(-\varrho \beta^2)}= \beta.
\label{asymmetricVpotentialstateseqn}
\ee
A plot of this result is shown in Figure~\ref{asymmetricVpotentialstatesplot}. There, it can be seen, for example, how the energy of the state $n$ varies as $\alpha_2$ changes, for $\alpha_1$ constant: as $\alpha_2\to 0$ ($\beta\to\infty$) the right branch of the potential goes to zero, and the bound states disappear; as $\alpha_2\to\infty$ ($\beta\to 0$) the right branch of the potential becomes an infinite barrier and the discrete bound states that ``survive'' are only those coming from the solutions of ${\rm Ai}(-\varrho)=0$ (those coming from ${\rm Ai}'(-\varrho)=0$, which are also poles of \eqref{finalgreenabsgreen}, are no longer solutions of the limit case).

\subsection{Half-oscillator-half-linear potential}

\begin{figure}[h]
\centering
\includegraphics[width=0.33\textwidth]{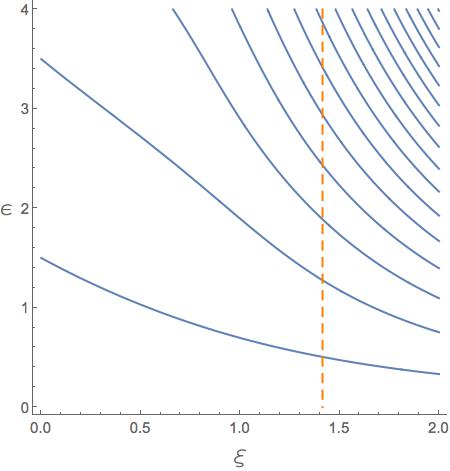} 
\caption{\small The first bound states of the asymmetric composite potential 
\eqref{composite_potential} from equation \eqref{tururu} as a function of $\xi=(2m\hbar\omega^3)^{1/6}/\alpha$, for $\hbar^2=2m$; the particular value $\xi=\sqrt{2}$ is stressed with a vertical dashed orange line. 
\label{figuray}}
\end{figure}

As an application of the results of the previous sections, let us consider an interesting example, the composite potential consisting of a harmonic oscillator for $x\leq 0$ and a constant force for $x\geq 0$:
\be
V(x)= \frac12 m\omega^2x^2\, \theta(-x)+ \alpha^3\, x\, \theta(x).
\label{composite_potential}
\ee 
In this case, the equation providing the bound states \eqref{(3.5)} turns out to be
\be
G_{ho}(0,0;E)+G_{\alpha}(0,0;E)=0,
\label{eqforbound_halfhalf}
\ee
where $G_{ho}(0,0;E)$ is given in \eqref{(8)} and $G_{\alpha}(0,0;E)$ in \eqref{finalgreenabsgreen}. Using the two parameters $\epsilon={E}/{\hbar\omega}$ and
$\xi= ({2m\hbar\omega^3})^{1/6}/{\alpha}=(2/(\mu\varphi))^{1/3}$, where $\mu,\varphi$ were already introduced in \eqref{physicalparameters},
equation \eqref{eqforbound_halfhalf} adopts the form
\be
\frac{  {\rm Ai}'(-\xi^2 \epsilon)  }{  \Gamma\left(\frac34-\frac{\epsilon}{2}\right)  } - \sqrt{2}\ \left(\frac{\hbar^2}{2m} \right)^{1/3} 
\xi \ \frac{{\rm Ai}(-\xi^2 \epsilon)  }{  \Gamma\left(\frac14-\frac{\epsilon}{2}\right)  }=0.
\label{tururu}
\ee
In Figure~\ref{figuray} a plot of the first bound states of the composite potential \eqref{composite_potential} as a function of $\xi$ is shown. If $\alpha\to 0$ (that is, $\xi\to\infty$) the potential becomes zero for $x\geq 0$ and then the bound states disappear completely. If $\alpha\to \infty$ (that is, $\xi\to 0$) an infinite barrier emerges for $x\geq 0$, and only the odd states of the harmonic potential survive.

\begin{figure}[h]
\centering
\includegraphics[width=0.33\textwidth]{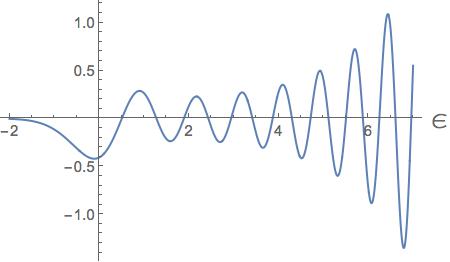} 
\qquad\qquad
\includegraphics[width=0.33\textwidth]{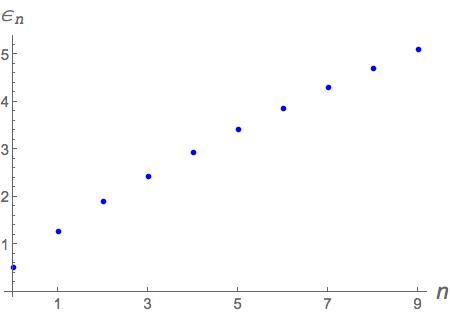} 
\caption{\small For the particular value $\xi=\sqrt{2}$, a plot of the LHS of equation \eqref{tururu} (with $\hbar^2=2m$) as a function of $\epsilon$, where the zeros are clearly visible,  is given on the left. On the right a plot of the numerical values of the zeros, given in Table~\ref{tab:title}.
\label{figuraz}}
\end{figure}

\begin {table}[h]
\begin{center}
\begin{tabular}{||c|c||c||c|c||}
\hline
$\epsilon_0$ &  0.50501  & &	$\epsilon_5$ &    3.41789\\
\hline
$\epsilon_1$ &  1.27615  & &	$\epsilon_6$ &    3.86844\\
\hline
$\epsilon_2$ &   1.88901 & &	$\epsilon_7$ &    4.29867\\
\hline
$\epsilon_3$ &   2.43392 & &	$\epsilon_8$ &    4.71332\\
\hline
$\epsilon_4$ &   2.94119 & &	$\epsilon_9$ &    5.11461 \\
\hline
\end{tabular}
\caption {First ten bound states energies for the composite potential \eqref{composite_potential}, evaluated numerically from \eqref{tururu} 
for the particular value $\xi=\sqrt{2}$.}
 \label{tab:title} 
\end{center}
\end {table}

For the special value $\xi=\sqrt{2}$, a plot of the left hand side of equation \eqref{tururu} as a function of $\epsilon$ is given in Figure~\ref{figuraz}. In this last plot the zeros of the function are clearly visible, and can be computed numerically. The calculation yields the results shown on Table~\ref{tab:title} for the first ten bound states, which are graphically represented on the RHS of Figure~\ref{figuraz}.

\subsection{Harmonic oscillator and symmetric linear potential}

Let us consider now the following potential, which is clearly related to the previous examples:
\be
V(x)= \frac12 m\omega^2x^2 + \alpha^3\, |x| .
\label{potentialx2x}
\ee 
Using the parameters $\epsilon$, $\mu$, $\varphi$ and $\sigma$ defined in \eqref{physicalparameters},
the wave function solution of the corresponding Schr\"odinger equation, which is continuous at $x=0$ and bounded at $\pm\infty$, can be expressed in terms of parabolic cylinder functions as
\be
\psi(x)=\left\{\begin{array}{ll}
\psi_1(x)=C\ D_{\sigma-1/2} \bigl(  \mu x+\mu \varphi \bigr) ,&  x\geq 0, \\ [2ex]
\psi_2(x)=C\ D_{\sigma-1/2} \bigl( - \mu x+\mu\varphi \bigr) ,&  x\leq 0.
\end{array}
\right.
\label{www}
\ee
The results of \cite{GGV} show us how to build the Green function for a problem like the previous one, gluing the solutions in two different regions $(a,0)$ and $(0,b)$. In particular, as in the present case, if we consider $a\to -\infty$ and $b\to +\infty$, the Green function, as a solution of an equation similar to \eqref{greentilde} or \eqref{partialdifeqnforgreenabsx}, is obtained taking the appropriate limits in equation (12) of \cite{GGV}, and we get
\be
{\widetilde G}(x,x';E)=\frac{\psi_1(x_>)\ \psi_2(x_<)}{W[\psi_1(x'),\psi_2(x')]},
\qquad
G(x,x';E)=-\frac{2m}{\hbar^2} \frac{\psi_1(x_>)\ \psi_2(x_<)}{W[\psi_1(x'),\psi_2(x')]},
\label{generalformofgreenfunctiongluing}
\ee
where the denominator in \eqref{generalformofgreenfunctiongluing} is the Wronskian of the functions $\psi_1(x')$ and $\psi_2(x')$. Using the two functions in \eqref{www}, and evaluating the Wronskian at $x'=0$, we get for the present problem
\be
G_{\mu,\varphi}(x,x';E)=\frac{2m}{\mu\hbar^2} \frac{D_{\sigma-1/2} \bigl( - \mu x_<+\mu \varphi \bigr)\ 
D_{\sigma-1/2} \bigl(  \mu x_>+\mu \varphi \bigr)}{  D_{\sigma-1/2}\bigl(\mu \varphi\bigr) \left[ \mu \varphi  D_{\sigma-1/2}\bigl(\mu \varphi\bigr)-2 D_{\sigma+1/2}\bigl(\mu \varphi\bigr)  \right] }.
\label{generalformofgreenfunctiongluingoscillatorlinear}
\ee
The poles of this Green function, that is, the zeros of the denominator, are given by
\be
D_{\sigma-1/2}\bigl(\mu \varphi\bigr) \left[ \mu \varphi  D_{\sigma-1/2}\bigl(\mu \varphi\bigr)-2 D_{\sigma+1/2}\bigl(\mu \varphi\bigr)  \right]=0.
\label{boundstatesx2absx}
\ee

\begin{figure}[h]
\centering
\includegraphics[width=0.33\textwidth]{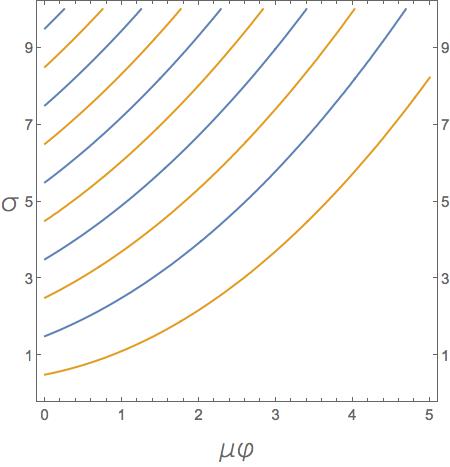} 
\qquad\qquad
\includegraphics[width=0.33\textwidth]{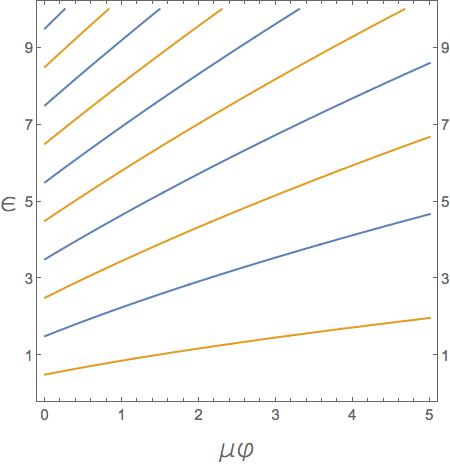} 
\caption{\small The bound states for the potential \eqref{potentialx2x} obtained from equation \eqref{boundstatesx2absx}: on the left in terms of the parameters $\sigma$ and $\mu \varphi$ given in \eqref{physicalparameters}; on the right the dimensionless energy $\epsilon=E/\hbar\omega$ as a function of $\mu \varphi$. The even states (orange curves) correspond to the solutions of $(\mu \varphi ) D_{\sigma-1/2}\bigl(\mu \varphi\bigr)=2 D_{\sigma+1/2}\bigl(\mu \varphi\bigr) $, and the odd states (blue curves) are the solutions of $D_{\sigma-1/2}\bigl(\mu \varphi\bigr)=0$.
\label{figurax2x}}
\end{figure}

It is possible to analyze the limit $\alpha,\varphi\to 0$ (which implies $\sigma\to E/\hbar\omega$) of this expression, which corresponds to the simple harmonic oscillator Green function \eqref{(8)}:
\be
G_{\mu,0}(x,x';E)=
\sqrt{\frac{\hbar}{4\pi m\omega}}  \Gamma\left( \frac12- \frac{E}{\hbar\omega}\right)\ 
 D_{\frac{E}{\hbar\omega}-\frac12} \bigl(  \mu x_> \bigr)\
 D_{\frac{E}{\hbar\omega}-\frac12} \bigl( - \mu x_< \bigr) = G_{ho}(x,x';E).
\label{generalformofgreenfunctionoscillator}
\ee

\section{Additional Dirac $\delta$ interaction potential}
The appearance of the Dirac delta function and its derivative  as potentials   in problems of the sort we consider has been examined in depth by Albeverio et al. \cite{albeverio}, whose work indicates that including them is compatible with the Green function techniques we have been using.
In 1987 Avakian et al. \cite{Avakian} introduced the singular oscillator model $V(x)=\frac{1}{4}\omega^2x^2+\Omega\delta(x-q)$, $q=0$, and discussed its bound states. We shall generalize their results by letting $q\ne0$:
\be
V_{a}(x,q)=a\, \delta(x-q), \quad a\in\mathbb R. 
\label{pointinteraction}
\ee
This generalization will be studied for the unperturbed potential $V_u(x)$ to be both the harmonic oscillator and the absolute value potential considered in the previous section.

We take into account the fact that the eigenfunction of eigenvalue $E$, corresponding to the potential 
$V(x)=V_u(x)+V_{a}(x,q)$, must satisfy the Lippman-Schwinger equation\footnote{In operator form we have $G_u(H_u-E)=I$ for general unperturbed Hamiltonian $H_u$ and Green function $G_u$. If we consider an eigenstate $|\psi\rangle$ of the perturbed Hamiltonian $H_u+V_a$ with eigenvalue $E$, then $(H_u+V_a-E)|\psi\rangle=0$. In addition, $G_u(H_u-E)|\psi\rangle=|\psi\rangle=- G_uV_a|\psi\rangle$. In the $x$-representation we have then \eqref{LS}.}
\be
\psi(x)=- \int_{-\infty}^{\infty}dx'\ G_u(x,x';E)\, V_{a}(x',q)\, \psi(x'),
\label{LS}
\ee
with $G_u(x,x';E)$ the ``unperturbed Green function'' given either by \eqref{(8)} or by \eqref{finalgreenabsgreen}.

\subsection{Harmonic oscillator with additional $\delta$ interaction}

Let us consider first the presence of a $\delta$-perturbation, like in \eqref{pointinteraction}, on the harmonic oscillator potential. Then, the Lippman-Schwinger equation \eqref{LS} is 
\be
\psi(x)=- a \int_{-\infty}^{\infty}dx'\ G_{ho}(x,x';E)\, \delta(x'-q)\, \psi(x')= - a \, G_{ho}(x,q;E)\,  \psi(q).
\label{LSoscillsolodelta}
\ee
Now, we set $x=q$ and find that the eigenvalues of the energy are given by $G_{ho}(q,q;E)=-1/a$, that is,
\be
\tau \, D_{\epsilon-\frac{1}{2}} \left(p\right)\,
D_{\epsilon-\frac{1}{2}} \left( -p\right)+\frac1{\Gamma\left(\frac{1}{2}-\epsilon \right)}=0,
\label{AP}
\ee
with $\epsilon=E/\hbar\omega$ and $p=\mu q$, where $\mu$ is given in \eqref{physicalparameters} and $\tau= a \sqrt{m/{\pi\omega\hbar^{3}}}$. Note that \eqref{AP} generalizes equation~(12) of \cite{Avakian}. Indeed, for $q=p=0$ our result completely agrees with that of Avakian et al. \cite{Avakian}. 

\begin{figure}[ht]
\centering
\includegraphics[width=0.33\textwidth]{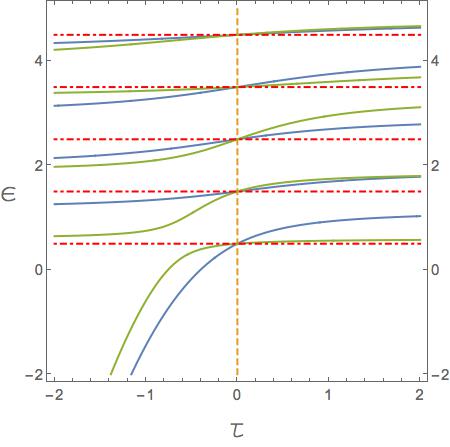}
\caption{\small First bound states of \eqref{AP} as a function of the intensity of the Dirac delta perturbation $\tau\propto a$, for two different positions of the singularity: $p=1/2$ (blue line) and $p=2$ (green line). The dotted red lines correspond to the energies of the unperturbed harmonic oscillator states, obtained also for $\tau=0$ (orange dashed line).}
\label{fig44}
\end{figure}

\begin{figure}[h]
\centering
\includegraphics[width=0.33\textwidth]{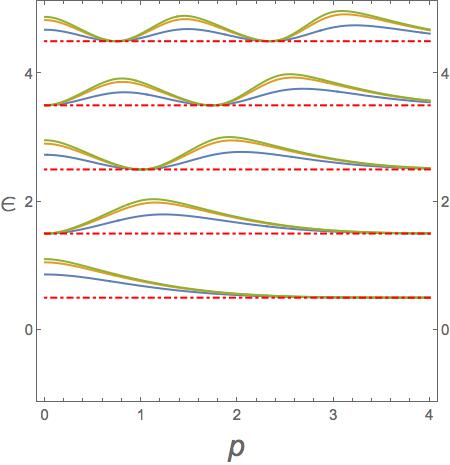}
\qquad\qquad
\includegraphics[width=0.33\textwidth]{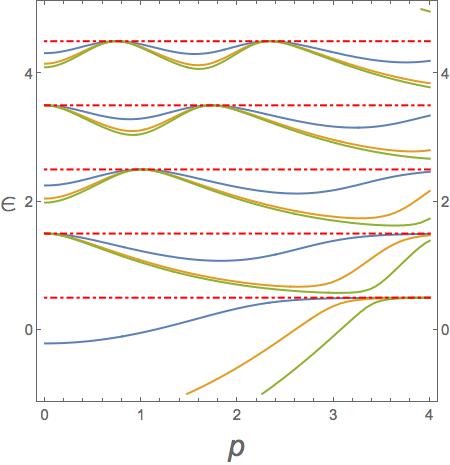}
\caption{\small Energy of the first bound states of \eqref{AP} as a function of the singularity coordinate $p=\mu q$, for several intensities $\tau\propto a$ of the Dirac delta perturbation. 
On the left we have three cases of ``repulsive'' delta-perturbations:: $\tau=0.5$, (blue line), $\tau=1$ (orange line) and $\tau=1.2$ (green line); on the right, three cases of attractive delta-perturbations: $\tau=-0.5$, (blue line), $\tau=-1$ (orange line) and $\tau=-1.2$ (green line).
In both figures the dotted red lines correspond to the energies of the unperturbed harmonic oscillator states, obtained also for $\tau=0$.
Observe that as $p$ grows, the perturbation on the lower states decreases. For either attractive or repulsive perturbation, at $p= 0$ the energy of the odd states do not change, but it does for the even states.}
\label{fig4554}
\end{figure}

The result we have obtained is completely symmetric in $q$, therefore, as in \cite{Avakian}, and without loss of generality, we will take in the sequel $q,p\geq 0$. In Figures~\ref{fig44} and \ref{fig4554} we show plots of the bound states of the oscillator modified by the presence of the Dirac delta perturbation $V_{a}(x,q)$. Note that if 
$a,\tau<0$ (attractive Dirac delta), the resulting bound state energies are lower than the unperturbed ones; on the contrary, if $a,\tau>0$ (repulsive Dirac delta), the new energies are higher than the unperturbed ones.

\subsection{Absolute value potential with additional $\delta$ interaction}

Let us analyze now the effect of the presence of a $\delta$ perturbation $V_{a}(x,q)=a\, \delta(x-q)$, $a\in\mathbb R$, on the eigenstates of the absolute value potential. This problem was considered in the past \cite{Xin}, but only for $q=0$. In the present case case, the Lippman-Schwinger equation \eqref{LS} is 
\be
\psi(x)=- a \int_{-\infty}^{\infty}dx'\ G_{\alpha}(x,x';E)\, \delta(x'-q)\, \psi(x')= - a\,  G_{\alpha}(x,q;E)\,  \psi(q),
\label{LSoscillsolodelta}
\ee
where $G_{\alpha}(x,x';E)$ is given in \eqref{finalgreenabsgreen}. Now, we set $x=q$ and find that the eigenvalues of the energy are given by $G_{\alpha}(q,q;E)=-1/a$, where $G_{\alpha}(x,x';E)$ in given in \eqref{finalgreenabsgreen}, that is, 
\be
\eta\,{\rm Ai}(\zeta q-\varrho)\,{\rm Ai}(-\zeta q-\varrho)={\rm Ai}(-\varrho)\,{\rm Ai}'(-\varrho),
\label{AP22}
\ee
where $\varrho$ and $\zeta$ are defined in \eqref{parameters_linearcase}, and 
$$
\eta=\frac{a}{2\alpha}  \left( \frac{2m}{\hbar^2} \right)^{2/3}.
$$
The complete Green function for this model can be obtained from the Dyson equation
\be
G(x,x';E)= G_{\alpha}(x,x';E)+ \int_{-\infty}^{\infty} G_{\alpha}(x,y;E)\, V_{a}(y,q)\, G(y,x';E)\ dy ,
\ee
and turns out to be
\be
G(x,x';E)= G_{\alpha}(x,x';E)+ a \ \frac{G_{\alpha}(x,q;E)\ G_{\alpha}(q,x';E)}{1-a \, G_{\alpha}(q,q;E)},
\ee
where $G_{\alpha}(x,x';E)$ is given in \eqref{finalgreenabsgreen}. 

In a similar way transcendental equations  can be obtained giving the variation of the energy levels of the harmonic oscillator plus linear potential \eqref{harmonicoscillator+Fx} and the harmonic oscillator plus symmetric linear potential
\eqref{potentialx2x} as a consequence of the addition of a Dirac delta perturbation as \eqref{pointinteraction}.

\section{Summary and discussion}

In this report we have applied the Green function technique to study a number of simple, but apparently new, quantum mechanical problems dealing with confining quadratic and linear potentials in one spatial dimension. For completeness  we have derived the Green functions involved, though at least one of them  has been available in the literature for many years \cite{Erdelyi}. Specifically, we have examined the harmonic oscillator in a constant force field, the asymmetric linear and quadratic oscillators,  two hybrids thereof and two cases of decoration by Dirac delta potentials. By means of a composite Green function matching technique, we have streamlined the formulation of the transcendental equations determining the bound-state energy levels and their dependence on various parameters. We have treated the decoration problem by means of the Lipmann-Schwinger bound-state formula which leads to the necessary eigenvalue equations more expeditiously than  by the wave function matching analysis used in prior studies \cite{Avakian,FR} of the decorated harmonic oscillator, confirming these pioneering calculations. We hope in the future to return to  the construction of the exact Green functions for the composite oscillators. This work also serves as an exploratory study of models and techniques which we hope to apply to a new class of inhomogeneous quantum wells.
Another natural step is to study stronger singular perturbations such as the Dirac delta derivatives considered in  \cite{FR,Alvarez,derivative,theorphys}. This analysis requires a more careful study and is  work in progress.

\section*{Acknowledgements}

Financial support is acknowledged to the Spanish MINECO (Project MTM2014-57129) and Junta de Castilla y
Le\'on (Project GR224). MLG thanks the Department of Theoretical Physics of the University of Valladolid for hospitality.


\begin{thebibliography}{99}
\bigskip

\bibitem{LEP}
 Physics at LEP, CERN 86-02, Vol.1 (1986).

\bibitem{Avakian} 
M.P. Avakian, G.S.  Pogosyan, A.N. Sissakian and V.M. Ter-Antonyan, Phys. Lett. A {\bf 124}, 233 (1987).

\bibitem{At}
 D.A. Atkinson and H.W. Crater, Amer. J. Phys. {\bf 43}, 301 (1975).
 
\bibitem{Vian}
 J. Viana-Gomes and N.M.R. Peres, Eur. J. Phys. {\bf 32}, 1377 (2011).

\bibitem{FI1}
 S. Fassari and G. Inglese, Helv. Phys. Acta {\bf 67}, 650 (1994).
 
 \bibitem{FI2}
 S. Fassari and G. Inglese, Helv. Phys. Acta {\bf 69}, 130 (1996).
 
 \bibitem{FI3}
 S. Fassari and G. Inglese, Helv. Phys. Acta {\bf 70}, 858 (1997).
 
 \bibitem{FR}
S. Fassari and F. Rinaldi, Rep. Math. Phys. {\bf 69}, 353 (2012).

\bibitem{albeverio} 
S. Albeverio,  F. Gesztesy, R, H$\phi$egh-Krohn, and H. Holden,  {\it Solvable Models in Quantum Mechanics}, AMS Publishers, Providence (2004).

\bibitem{Bel}
 M. Belloni and R.W. Robinett, Phys. Rep. {\bf 540}, 25 (2014).

\bibitem{Denev}
S. Denev, {\it Luminescent effects  in Quantum Well structures in Magnetic and Electric Fields}, Doctoral Thesis, University of Pittburgh (2005).

\bibitem{Barrio}
D.P. Barrio, M.L. Glasser, V.R. Velasco and F. Garcia-Moliner, J. Phys. Cond. Matter {\bf 1}, 4339 (1989).

\bibitem{Porubaev}
F.V. Porubaev and L.E. Golub, Phys. Rev. B {\bf 90}, 085314 (2014).
 
\bibitem{comp} 
M.L. Glasser, Amer. J. Phys. {\bf 47}, 739 (1979). 
 
\bibitem{Feynman}
R.P. Feynman and A.R. Hibbs, {\it Quantum Mechanics and Path Integrals: Emended Edition} (Dover, New York 2010).

\bibitem{Manoukian}
E.B. Manoukian, {\it Quantum Theory} (Springer, New York 2006).

\bibitem{Titchmarsh} 
E.C. Titchmarsh, {\it Eigenfunction Expansion associated with Second-order Differential Equations. Part One} (Oxford University Press, Oxford 1962). 

\bibitem{Khrebtukov} 
D.B. Khrebtukov and J.H. Macek, J. Phys. A {\bf 31}, 2853 (1998).

\bibitem{Erdelyi} 
A. Erdelyi et al. (Eds.), {\it The Bateman Project: Higher Transcendental Functions, Vol.II} (McGraw-Hill Book Company, New York 1953).

\bibitem{Buchholtz} 
M.L. Glasser,  Int. Trans. and Spec. Functions (to appear); arXiv:1502.00102 [math.CA].

\bibitem{GGV} 
M.L. Glasser, F. Garc\'ia-Moliner, V.R. Velasco, J. Appl. Phys. {\bf 68}, 4319 (1990).

\bibitem{NIST} 
F.~W.~J. Olver and D.~W. Lozier and R.~F. Boisvert and C.~W. Clark, {\it NIST Handbook of Mathematical Functions} (Cambridge University Press", New York 2010). Print companion to \url{http://dlmf.nist.gov/}.

\bibitem{Xin}
Wang Xin, Tang Liang-Hui, Wu Reng-Lai, Wang Nan and Liu Quan-Hui, Commun. Theor. Phys. {\bf 53}, 247 (2010).

\bibitem{Alvarez} 
J.J. \'Alvarez, M. Gadella, M.L. Glasser, L.P. Lara and L.M. Nieto, J. Phys.: Conf. Ser. {\bf 284}, 012009 (2011).

\bibitem{derivative} 
M. Gadella, J. Negro and L.M. Nieto, Phys. Lett. A {\bf 373}, 1310 (2009). 

\bibitem{theorphys} 
M. Gadella,  M.L. Glasser and L.M. Nieto,  Int. J. Theor. Phys.  {\bf 50}, 2144 (2011).




\end{thebibliography}
\end{document}